\documentclass[]{aa}

\usepackage{psfig}

\usepackage{amsmath,amssymb,graphicx}

\usepackage{natbib}

\bibpunct{(}{)}{;}{a}{}{,}

\begin{document}

\headnote{Research Note}

\title{Direct measurements of black hole charge with future astrometrical missions}

\author{A.F. Zakharov\inst{1,2,3,4}, F. De Paolis\inst{5}, G. Ingrosso\inst{5}, A.A. Nucita\inst{5}}


\offprints{Zakharov A.F., \email{zakharov@itep.ru}}

\institute{ National Astronomical Observatories of Chinese Academy
of Sciences, 20A Datun Road, Chaoyang District, Beijing 100012,
China \and
 Institute of Theoretical and Experimental Physics,
           25, B.Cheremushkinskaya st., Moscow, 117259, Russia,
\and Astro Space Centre of Lebedev Physics Institute, 84/32,
Profsoyuznaya st.,
             Moscow, 117810, Russia,
             \and Joint Institute for Nuclear Research, Dubna,
             Russia
\and Department of Physics, University of Lecce and INFN, Section
of Lecce, Via Arnesano, I-73100 Lecce, Italy }

\date{Received / accepted }

\abstract{Recently, \cite{ZNDI05} considered the possibility of
evaluating the spin parameter and the inclination angle for Kerr
black holes in nearby galactic centers by using future advanced
astrometrical instruments. A similar approach which uses the
characteristic properties of gravitational retro-lensing images
can be followed to  measure the charge of a Reissner-Nordstr\"{o}m
black hole. Indeed, in spite of the fact that their formation
might be problematic, charged black holes are objects of intensive
investigations. From a theoretical point of view, it is well-known
that a black hole is described by only three parameters, namely,
its mass $M$,  angular momentum $J$, and charge $Q$. Therefore, it
would be important to have a method for measuring all these
parameters, preferably by independent model of any. In this paper,
we propose a procedure to measure the black hole charge by using
the size of the retro-lensing images that can be revealed by
future astrometrical missions. A discussion of the Kerr-Newmann
black hole case is also offered.

 }

 \authorrunning{A.F. Zakharov et al.}

 \titlerunning{Measurements of Black Hole Charge}

\maketitle

\section{Introduction}

"Black holes have no hair" means that a black hole is characterized
by only three parameters ("hairs"), its mass $M$, angular momentum
$J$, and  charge $Q$ (see, e.g. \citealt{MTW,Wald84} or
\citealt{Heusler98} for a more recent review). Therefore, in
principle, charged black holes can be formed, although astrophysical
conditions that lead to their formation may look rather
problematic: 
see, for example,
\cite{Zamir93,Ruffini99,Ruffini00,Lee01,Perjes03,Ray03,Moderski04,Vogt04,Lemos04,Sereno04,Ghezzi05}.
Nevertheless, one  can not claim that their existence is forbidden
by theoretical or observational arguments.

Charged black holes are also  objects of intensive studies, since
they are described by Reissner-Nordstr\"{o}m geometry which is  a
static, spherically
 symmetrical  solution of Yang-Mills-Einstein equations with fairly natural
 requirements on asymptotic behavior of the solutions \citep{Gal1,Gal2,Gal3,Lee}. The
 Reissner - Nordstr\"{o}m metric thus describes a spherically symmetric
 black hole with a color charge and (or) a magnetic monopole (see also \citealt{Heusler98}).

The formation of retro-lensing images (also known as mirage,
shadows, or "faces" in the literature) due to the strong
gravitational field effects near black holes has been investigated
by several authors
\citep{Holz02,Geralico03,Geralico04,ZNDI05,ZNDI05b,Zakharov_Protvino04}.
The question that naturally arises is whether these images are
observable or not. It has been shown that the retro-lensing image
around the black hole at the Galactic Center (Sgr $A^*$) due to
$S_2$ star is observable in the K-band (peaked at 2.2 $\mu$m) by
the next generation infra-red space-based missions. The effects of
retro-lensing image shapes due to black hole spin has also been
investigated \citep{Geralico04,ZNDI05,Zakharov_Protvino04,Zakharov_Texas04}.

In this paper we focus on the possibility of measuring the black
hole charge as well, and we present an analytical dependence of
mirage size on the black hole charge. Indeed, future space missions
like Radioastron in radio band or MAXIM in X-ray band have angular
resolution close to the shadow size for massive black holes in the
center of our own and nearby galaxies.

\section{Basic definitions and equations}

The expression for the Reissner - Nordstr\"{o}m metric in natural
units ($G=c=1$) has the form
\begin {eqnarray}
  ds^{2}=-(1-\frac{2M}{r}+\frac{Q^{2}}{r^{2}})dt^{2}+(1-\frac{2M}{r}+\frac{Q^{2}}{r^{2}})^{-1}dr^{2}+
  \nonumber \\
+r^{2}(d{\theta}^{2}+{sin}^{2}\theta d{\phi}^{2}).
\label{RN_Lecce_0}
\end {eqnarray}


Applying the Hamilton-Jacobi method to the problem of 
geodesics in the Reissner - Nordstr\"{o}m metric, the motion of a
test particle in the $r$-coordinate can be described by following
equation (see, for example, \citealt{MTW})

\begin {eqnarray}
    r^{4}(dr/d\lambda)^{2}=R(r),\label{RN_Lecce_1}
\end {eqnarray}
where
\begin {eqnarray}
&&  R(r)=P^{2}(r)-\Delta({\mu}^{2}r^{2}+L^{2}), \nonumber\\
&&  P(r)=Er^{2}-eQr,    \label{RN_Lecce_2}\\
&&  \Delta=r^{2}-2Mr+Q^{2}. \nonumber
\end {eqnarray}
Here, the constants $\mu, E, L$, and $e$ are associated with the
particle; i.e. $\mu$ is its mass, $E$ is energy at infinity, $L$
is its angular momentum at infinity, and $e$ is the particle's
charge.

    We shall consider the motion of uncharged particles $(e=0)$ below.
In this case, the expression for the polynomial $R(r)$ takes the
form
\begin {eqnarray}
R(r)=(E^{2}-{\mu}^{2})r^{4}+
2M{\mu}r^{3}-(Q^{2}{\mu}^{2}+L^{2})r^{2}+\nonumber\\
+2ML^{2}r-Q^{2}L^{2}. \label{RN_Lecce_3}
\end {eqnarray}

     Depending on the multiplicities of the roots of the
polynomial $R(r)$,  we can have three types of motion in the $r$ -
coordinate \citep{Zakh1}. In particular, by defining
$r_{+}=1+\sqrt{1-Q^{2}}$, we have:

 (1) if the polynomial $R(r)$
has no roots  for $r\geq r_{+}$,  a test particle is captured by
the black hole;

(2) if $R(r)$ has roots  and
$(\partial{R}/\partial{r})(r_{max})\neq 0$  with $r_{max} > r_+$
($r_{max}$ is the maximal root), a particle is scattered after
approaching the black hole;

(3) if  $R(r)$ has a root  and $R(r_{max}) = (\partial{R}/
\partial{r})(r_{max})=0$,
the particle now takes an infinite proper time to approach the
surface $r = const$.

    If we are considering a photon ($\mu = 0$), its motion in the $r$-coordinate depends
on the root multiplicity of the polynomial $\hat{R}(\hat{r})$
\begin {eqnarray}
\hat{R}(\hat{r})={R(r)}/({M^{4}E^{2}})={\hat{r}}^{4}-\xi^{2}{\hat{r}}^{2}+2{\xi}^{2}\hat{r}
 -{\hat{Q}}^{2}{\xi}^{2},\label{RN_Lecce_4}
\end {eqnarray}
where $\hat {r}=r/M, \xi=L/(Me)$ and $\hat {Q}=Q/M.$

One can see from Eq. (\ref{RN_Lecce_4}) and Eqs.
(\ref{RN_Lecce_2})  that the black hole charge may substantially
influence the photon motion at small radii ($r \approx 1$), while
the charge effect is almost negligible at large radial coordinates
of photon trajectories ($r
>> 1$). In the last case, we should keep in mind that the charge may cause
only small corrections to photon motion.

\section{Capture cross section of photons by a Reissner -- Nordstr\"{o}m black hole}

    Let us consider the problem of the capture cross section of
 a photon by a charged black hole. It is clear that the critical value of the
 impact parameter for a photon to be captured by a Reissner - Nordstr\"{o}m black hole
depends on the multiplicity root condition of the polynomial
$R(r)$, i. e. the condition for a vanishing discriminant
\citep{Zakh3,Zakh,Zakh2,Zakh94}. In particular,  it was shown that
the vanishing discriminant condition approach is more powerful
than the procedure that excludes $r_{max}$ from the following
system
\begin {eqnarray}
R(r_{max}) = 0, \quad 
\quad \dfrac{\partial{R}}{\partial{r}}(r_{max})=0,
\label{RN_Lecce_4a}
\end {eqnarray}
as  was done, for example, by \cite{chandra} to solve similar
problems.

Introducing the notation $\xi^{2}=l, Q^{2}=q$, we obtain
\begin {eqnarray}
    R(r)=r^{4}-lr^{2}+2lr-qr.\label{RN_Lecce_5}
\end {eqnarray}
The discriminant $\Delta$ of the polynomial $R(r)$ has the form,
as it was shown by \cite{Zakh,Zakh2,Zakh94}:
\begin {eqnarray}
   \Delta
=16l^{3}[l^{2}(1-q)+l(-8q^{2}+36q-27)-16q^{3}]. \label{RN_Lecce_6}
\end {eqnarray}
The polynomial $R(r)$ thus has a multiple root  if and only if
\begin {eqnarray}
  l^{3}[l^{2}(1-q)+l(-8q^{2}+36q-27)-16q^{3}]=0. \label{RN_Lecce_7}
\end {eqnarray}
Excluding the case $l=0$, which corresponds to a multiple root at
$r=0$, we find
 that the polynomial $R(r)$ has a multiple root for $ r\geq r_{+}$ if and only if
\begin {eqnarray}
 l^{2}(1-q)+l(-8q^{2}+36q-27)-16q^{3}=0. \label{RN_Lecce_8}
\end {eqnarray}
If $q=0$, we obtain the well-known result for a Schwarzschild
black hole \citep{MTW,Wald84,Lightman}, $l=27$, or
$L_{cr}=3\sqrt{3}$. If $q=1$, then $l = 16$,  or $L_{cr}=4$, which
also  corresponds to numerical results given by
\cite{Young,de_Vries00,Takahashi05}.

   The photon capture cross section for an  extremely charged
    black hole  turns out to be considerably smaller than the capture cross section of a
 Schwarzschild black hole. The critical value of the impact parameter,
 characterizing the capture cross section for a Reissner - Nordstr\"{o}m
black hole, is determined by the equation
\citep{Zakh,Zakh2,Zakh94}
\begin {eqnarray}
l=\frac{b_1+\sqrt{{b_1}^2+64q^{3}(1-q)}}{2(1-q)}.
\label{RN_Lecce_9}
\end {eqnarray}
where $b_1=8q^{2}-36q+27$.

 Substituting Eq.(\ref{RN_Lecce_9}) into the
expression for the coefficients of the polynomial $R(r)$, it is
easy to calculate the radius of the unstable circular photon
orbit, which is the same as
 the minimum periastron distance. The orbit of a
 photon moving from infinity with the critical impact parameter, determined
in accordance with Eq.(\ref{RN_Lecce_9}), spirals into  circular
orbit.

As  was explained by \cite{ZNDI05,ZNDI05b,Zakharov_Protvino04}
this leads to the formation of shadows  described by the critical
value of $L_{cr}$; or in other words, in the spherically symmetric
case, shadows are circles with radii $L_{cr}$. Therefore, by
measuring the shadow size, one could evaluate the black hole
charge in  black hole mass units $M$.

\begin{figure}[th!]
\begin{center}
\includegraphics[width=8.5cm]{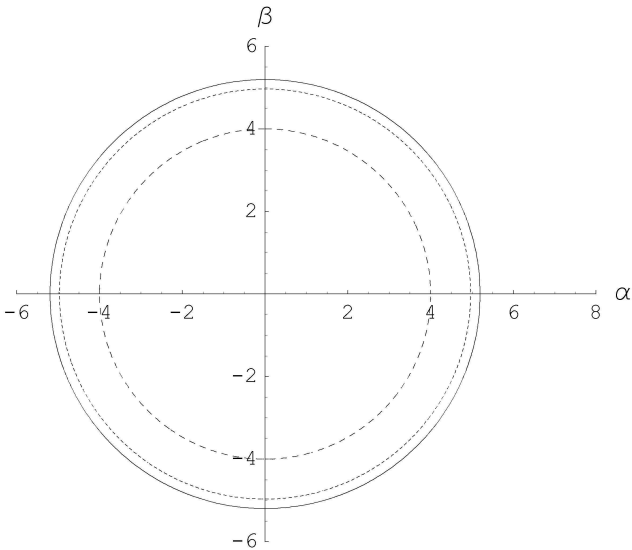}
\end{center}
\caption{Shadow (mirage) sizes are shown for selected charges of
black holes $Q=0$ (solid line), $Q=0.5$ (short dashed line), and
$Q=1$ (long dashed line).}
 \label{Fig1}
\end{figure}

\begin{figure}[th!]
\begin{center}
\includegraphics[width=9cm]{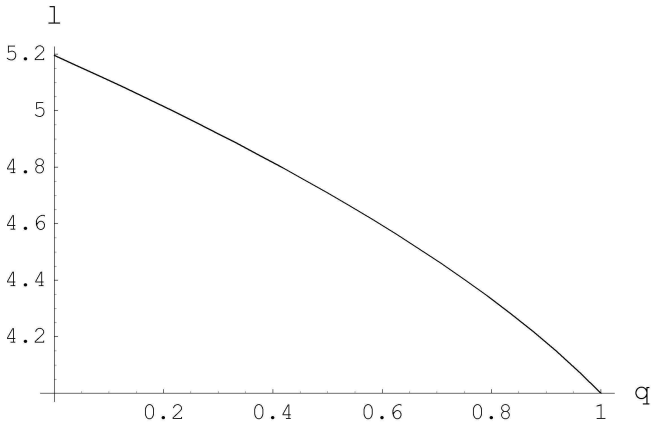}
\end{center}
\caption{The mirage radius $l$ 
is shown as a
function of the black hole charge $q$ ($l$ and $q$ are given in
units of $M$).} \label{Fig2}
\end{figure}

\section{The space RADIOASTRON interferometer}

The space-based radio telescope RADIOASTRON \footnote{See web-site
http://www.asc.rssi.ru/radioastron/ for more information.} is
planned to be launched within few next years. \footnote{This
project was proposed by the Astro Space Center (ASC) of Lebedev
Physical Institute of the Russian Academy of Sciences (RAS) in
collaboration with other institutions of RAS and RosAviaKosmos.
Scientists from 20 countries are developing the scientific payload
for the satellite by providing by ground-based support to the
mission.} This space-based 10-meter radio telescope will be used
for space -- ground VLBI observations. The measurements will have
extremely high angular resolutions, namely about 1 -- 10 $\mu as $
(in particular about 8 $\mu as $ at the shortest wavelength of
1.35 cm and a standard orbit,\footnote{The satellite  orbit  will
have high apogee, and its rotation period around Earth will be 9.5
days, which evolves as a result of the weak gravitational
perturbations from the Moon and the Sun. The perigee has been
planned to be between $10^4$ and $7\times 10^4$ km and the apogee
between 310 and 390 thousand kilometers. The basic orbit
parameters will be the following: the orbital period is P = 9.5
days, the semi-major axis is a = 189 000 km, the eccentricity is e
= 0.853, the perigee is H = 29 000 km.} and could be about 0.9
$\mu as $ for the high orbit configuration at the same wavelength.
Four wave bands will be used corresponding to $\lambda=1.35$~cm,
$\lambda=6.2$~cm, $\lambda=18$~cm, $\lambda=92$~cm (see Table 1).
A detailed calculation of the high-apogee evolving orbits ($B_{\rm
max}$) can be done, once the exact launch time is known.

After several years of observations,  it should be possible to
move the spacecraft  to a much higher orbit (with apogee radius
about 3.2 million km), by additional spacecraft maneuvering using
the gravitational force of the Moon.  The fringe sizes (in $\mu as
$) for the apogee of the above-mentioned orbit and for all
RADIOASTRON wavelengths are given in Table 1.

\begin{table}
\begin{center}
\caption[]{The fringe sizes (in micro arc seconds) for the
standard and advanced apogees $B_{max}$ (350 000 and 3 200 000~km,
respectively).}

\begin{tabular}{|c|c|c|c|c|}
\hline
$B_{max}({\rm km}) \backslash \lambda ({\rm cm})$ & 92 & 18 & 6.2 & 1.35 \\
\hline \hline
$3.5\times 10^{5}$ & 540 & 106 & 37 & 8 \\
\hline
$3.2 \times 10^{6}$ & 59 & 12 & 4 & 0.9 \\
\hline
\end{tabular}
\end{center}
\label{tabl1}
\end{table}
By comparing Fig. 1,2 and Table 1, one can see that there are
non-negligible chances to observe such mirages around the black hole
at the Galactic Center and in nearby AGNs and microquasars in the
radio-band using RADIOASTRON facilities.

We also mention that this high resolution in radio band will be
achieved also by Japanese VLBI project VERA (VLBI Exploration of
Radio Astrometry), since the angular resolution aimed at will be
at the 10 $\mu as$ level \citep{Sawada00,Honma02}. Therefore, the
only problem left is to have a powerful enough radio source to
illuminate a black hole in order to have retro-lensing images
detectable by such radio VLBI telescopes as RADIOASTRON or VERA.

\section{Searches for mirages near Sgr $A^*$ with RADIOASTRON}

 Radio, near-infrared,
and X-ray spectral band observations are developing very rapidly
\citep{Lo98,Lo99,Genzel03,Ghez04,Baganoff01,Baganoff03,Bower02,Bower03,Narayan03,Bower04},\footnote{An
interesting idea to use radio pulsars to investigate the region
nearby black hole horizon was proposed recently by
\cite{Pfahl03}.} and it is known that Sgr $A^*$  harbors the
closest massive black hole with  mass estimated to be $4.07\times
10^6 M_{\odot}$ \citep{Bower04,Melia01,Ghez03,Schodel03}.

Following  the idea of \cite{Falcke00} and of
\cite{ZNDI05,ZNDI05b,Zakharov_Protvino04,ZNDI05d} we propose to use
the VLBI technique to observe mirages around massive black holes
and, in particular, towards the black hole at Galactic Center. To
evaluate the shadow shape \cite{Falcke00} used the ray-tracing
technique. The boundaries of the shadows are black hole mirages. We
use the length parameter $r_g=\dfrac{GM}{c^2}=6 \times 10^{11}$ cm
for the black hole at Sgr $A^*$ and an analytical approach (see
Sect. 2, 3) to calculate shadow sizes, as explained in the text. By
taking the distance of Sgr $A^*$ to be $D_{\rm GC}=8$~kpc, the
length $r_g$ corresponds to angular size $\sim 5~\mu  as$. Since the
minimum arc size for the considered mirages are at least $8~r_g$,
the standard RADIOASTRON resolution of about $8~\mu as$ is
comparable to the required precision. The resolution in the case of
the highest orbit and shortest wavelength is  $\sim 1~\mu as$ (see
Table 1), good enough to reconstruct the mirage shapes. As can be
seen from Fig. 1 and Table 1, it is clear that, in principle, it is
possible to evaluate the black hole charge $Q$ by observing the
shadow size. The mirage size difference between the extreme charged
black hole and Schwarzschild black hole case is about 30\% (the
mirage diameter for Schwarzschild black hole is about 10.4, and for
the extreme charged black hole, the diameter is equal to 8 or in
black hole mass units) and typical angular sizes are about $\sim
52~\mu as$ for the Schwarzschild and $\sim 40~\mu as$ for the
Reissner-Nordstr\"{o}m black hole cases, respectively. Therefore,
for Sgr $A^*$ a charged black hole could be distinguished by a
Schwarzschild black hole with RADIOASTRON, at least if its charge is
close to the maximal value. For stellar mass black holes, we need a
much higher angular resolution  to distinguish charged and uncharged
black holes, since the typical shadow (mirage) angular sizes are
about $2\times 10^{-5} \mu as$, even for galactic black holes.

Actually, from the mathematical point of view we proved the
following statement. For any positions of source and observer,
there is an emitted photon passing close enough to any point of
the mirage that can be caught by the observer. Thus, the mirage is
formed by an envelope of a family of photon trajectories or, in
other words, by a caustic surface.

Mirage formation under realistic assumptions for a specific case
was discussed by \cite{Falcke00}. However, instead of solving the
shadow formation problem by numerical simulations, one can use the
approach  developed by \cite{zakharov1,zakharov5,
zak_rep1,zak_rep2,zak_rep02a,zak_rep03,zak_rep03c,zak_rep03d,ZR_Nuovo_Cim03,ZMB04,Zak_SPIG04,ZR_5SCSLSA,ZR_NA_05,Zakharov_IJMPA_05}.
Mirages are boundaries of bright images, and  shadows are also
generated  for accretion disks. However,  for  thick accretion
disks as well, some parts of mirages could be observable, at least
in principle.

\section{Conclusions}

The angular resolution of the space RADIOASTRON interferometer will
be high enough to resolve radio images around black holes. By
measuring the mirage shapes one should be able to evaluate the black
hole mass, inclination angle (e.g. the angle between the black hole
spin axis and line of sight), and spin, if the black hole distance
is known. For example, for the black hole at the Galactic Center,
the mirage size is $\sim 52~\mu as$ for the Schwarzschild case.  In
the case of a Kerr black hole \citep{ZNDI05,ZNDI05b,Geralico04}, the
mirage is deformed depending on the black hole spin $a$ and on the
angle of the line of sight, but its size is almost the same. In the
case of a Reissner-Nordstr\"{o}m black hole, its charge changes the
size of the shadows up to 30 \% for the extreme charge case.
Therefore, the charge of the black hole can be measured by observing
the shadow size, if the other black hole parameters are known with
sufficient precision. In general, one could say that a measure of
the mirage shape (in size) allows to evaluate all the black hole
``hairs'' to be evaluated.

However, there are two kinds of difficulties when measuring mirage
shapes around black holes. First, the brightness of these images
or their parts (arcs) may not be sufficient for being detected by
RADIOASTRON. But, numerical simulations by \cite{Falcke00} and
\cite{Melia01} give hope that the shadow brightness could be high
enough to be detectable. Second, turbulent plasma effects near the
black hole horizon could give essential broadening of observed
images \citep{Bower04} leading to a confusion of the shadow image.

Recent observations of simultaneous X-ray and radio flares at 3
mm, 7 mm, 1.3 cm, and 2 cm with few-hundred second rise/fall
timescales gave indirect evidence that X-ray and radio radiation
from the close vicinity of Sgr A$^*$ has been detected
\citep{Baganoff01}.

A few years ago the possibility of observing the images of distant
sources around black holes in the X-ray band was discussed by
\cite{White00} and \cite{Cash00} by using X-ray interferometer.
Indeed, the aim of the MAXIM project is to realize a space-based
X-ray interferometer capable of observing with angular resolution
as small as $0.1\mu as$.

One could also mention that if the emitting region has a
degenerate position with respect to the line of sight (for
example, the inclination angle of an accretion disk is $\gtrsim
85^0$) strong bending effects do appear (see, \citealt{Matt93},
\citealt{zak_rep03b}).

In spite of the difficulties of measuring the shapes of shadow
images, to look at black hole ``faces'' is an attractive challenge
since mirages outline the ``faces'' and correspond to a fully
general relativistic description of the region near the black hole
horizon without any assumption about a specific model for
astrophysical processes around black holes; of course, we assume
that there are sources illuminating black hole surroundings. There
is no doubt that the rapid growth of observational facilities will
give a chance to measure the mirage shapes using not only
RADIOASTRON facilities but also other instruments and spectral
bands, like the X-ray interferometer MAXIM,  the RADIOASTRON
mission, or other space-based interferometers in millimeter and
sub-millimeter bands \citep{Miyoshi04}.

\begin{acknowledgements}
    AFZ would
like to thank the Dipartimento di Fisica Universit\`a di Lecce,
INFN, Sezione di Lecce and National Astronomical Observatories of
Chinese Academy of Sciences, Beijing for their hospitality. AFZ
thanks also the National Natural Science Foundation of China (NNSFC)
for partial financial support of the work (Project \# 10233050).

\end{acknowledgements}

\end{document}